\documentclass[aps,prb,preprint,showpacs,superscriptaddress]{revtex4}

\usepackage{amsmath}
\usepackage{amssymb}
\usepackage{graphicx}

\begin{document}
\bibliographystyle{apsrev}

\title{Hysteretic jumps in the response of layered superconductors
to electromagnetic fields}

\author{V.A.~Yampol'skii}
\affiliation{Advanced Science Institute, The Institute of Physical
and Chemical Research (RIKEN), Wako-shi, Saitama, 351-0198, Japan}
\affiliation{ A.Ya.~Usikov Institute for Radiophysics and
Electronics Ukrainian Academy of Sciences, 61085 Kharkov, Ukraine}

\author{T.M.~Slipchenko}
\affiliation{ A.Ya.~Usikov Institute for Radiophysics and
Electronics Ukrainian Academy of Sciences, 61085 Kharkov, Ukraine}
\affiliation{V.N.~Karazin Kharkov National University, 61077
Kharkov, Ukraine}

\author{Z.A.~Mayzelis}
\affiliation{ A.Ya. Usikov Institute for Radiophysics and
Electronics Ukrainian Academy of Sciences, 61085 Kharkov, Ukraine}
\affiliation{V.N.~Karazin Kharkov National University, 61077
Kharkov, Ukraine}

\author{D.V.~Kadygrob}
\affiliation{V.N.~Karazin Kharkov National University, 61077
Kharkov, Ukraine}

\author{S.S.~Apostolov}
\affiliation{ A.Ya.~Usikov Institute for Radiophysics and
Electronics Ukrainian Academy of Sciences, 61085 Kharkov, Ukraine}
\affiliation{V.N.~Karazin Kharkov National University, 61077
Kharkov, Ukraine}

\author{S.E.~Savel'ev }
\affiliation{Advanced Science Institute, The Institute of Physical
and Chemical Research (RIKEN), Wako-shi, Saitama, 351-0198, Japan}
\affiliation{Department of Physics, Loughborough University,
Loughborough LE11 3TU, UK}

\author{Franco Nori}
\affiliation{Advanced Science Institute, The Institute of Physical
and Chemical Research (RIKEN), Wako-shi, Saitama, 351-0198, Japan}
\affiliation{Department of Physics, Center for Theoretical
Physics, Applied Physics Program, Center for the Study of Complex
Systems, University of Michigan, Ann Arbor, MI 48109-1040, USA}

\begin{abstract}
We consider here a layered superconductor subject to an externally
applied moderately-strong electromagnetic field. We predict
hysteretic jumps in the dependence of the surface reactance of the
superconductor on the amplitude $H_0$ of the incident
electromagnetic wave. This very unusual nonlinear phenomenon can
be observed in thin superconducting slabs at not very strong ac
amplitudes, if the frequency of the irradiating field is close to
the Josephson plasma frequency. Using the set of coupled
sine-Gordon equations, we derive the expression for the phase
shift $\chi$ of the reflected wave and obtain the conditions for
the appearance of hysteresis in the $\chi(H_0)$-dependence.
\end{abstract}

\date{\today}

\maketitle

%-------------------------------------------------------------------------------------
%-------------------------------------------------------------------------------------
\section{Introduction}

There has been a recent surge of studies of electromagnetic waves
(EMWs) propagating in artificially fabricated media (see, e.g.,
Refs.~\onlinecite{nature-rev,pw,ptd}), including metals with
modulated properties~\cite{metal}, arrays of coupled
waveguides~\cite{wg}, left-hand materials~\cite{lh,kiv,bliokh},
and layered superconductors~\cite{surf}. The excitation of these
waves can produce a large variety of resonance
anomalies~\cite{anomaly} in the reflectivity, transmissivity, and
absorptivity, offering new types of optical nano-devices.

The recent increase of these type of studies is related to
nonlinear surface and waveguide EM modes (see, e.g.,
Refs.~\onlinecite{nature-rev,wg,kiv}). In this broad context, a
layered superconductor is a medium favoring the propagation of
nonlinear~\cite{natphys,waveGad,n-lin} and surface~\cite{surf}
waves in the (important for applications~\cite{applic,tonouchi})
terahertz (THz) and sub-THz frequency ranges. Both the existence
of surface waves~\cite{surf} and the nonlinear
effects~\cite{natphys,waveGad,n-lin} occur due to the gap
structure~(see, e.g., Ref.~\onlinecite{mish}) of the spectrum of
Josephson plasma waves, which was experimentally observed via
Josephson plasma resonance~\cite{firstJPR}. The nonlinearity of
Josephson plasma waves with frequency $\omega$ close to the
Josephson plasma frequency $\omega_J$ becomes important even at
small field amplitudes $\propto |1-\omega^2/\omega_J^2|^{1/2}$. In
close analogy to nonlinear optics~\cite{books}, the nonlinear JPWs
exhibit numerous remarkable features~\cite{natphys,waveGad,n-lin},
including the slowing down of light, self-focusing effects, and
the pumping of weaker waves by stronger ones. However, the
nonlinearity of EMWs in layered superconductors is quite different
from optical nonlinearities. This leads one to expect very unusual
phenomena in the EMW propagation in this nonlinear media.

In this paper, we predict and analyze theoretically one of such
unexpected nonlinear effects in a thin slab of a layered
superconductor subject to an externally applied electromagnetic
wave. We show that, under specific conditions, the amplitude
dependence of the phase $\chi$ of the reflected wave becomes
many-valued. This should result in hysteretic jumps of $\chi$ when
sweeping the amplitude of the incident wave, a phenomenon which is
very unusual for conductors and superconductors.

\section{Problem statement and equations for the electromagnetic field}

\begin{figure}[!htp]
\begin{center}
%\hspace*{-0.9cm} \vspace*{-0.9cm}
%\includegraphics*[width=8cm]{fig1.eps}
\includegraphics*[width=17cm]{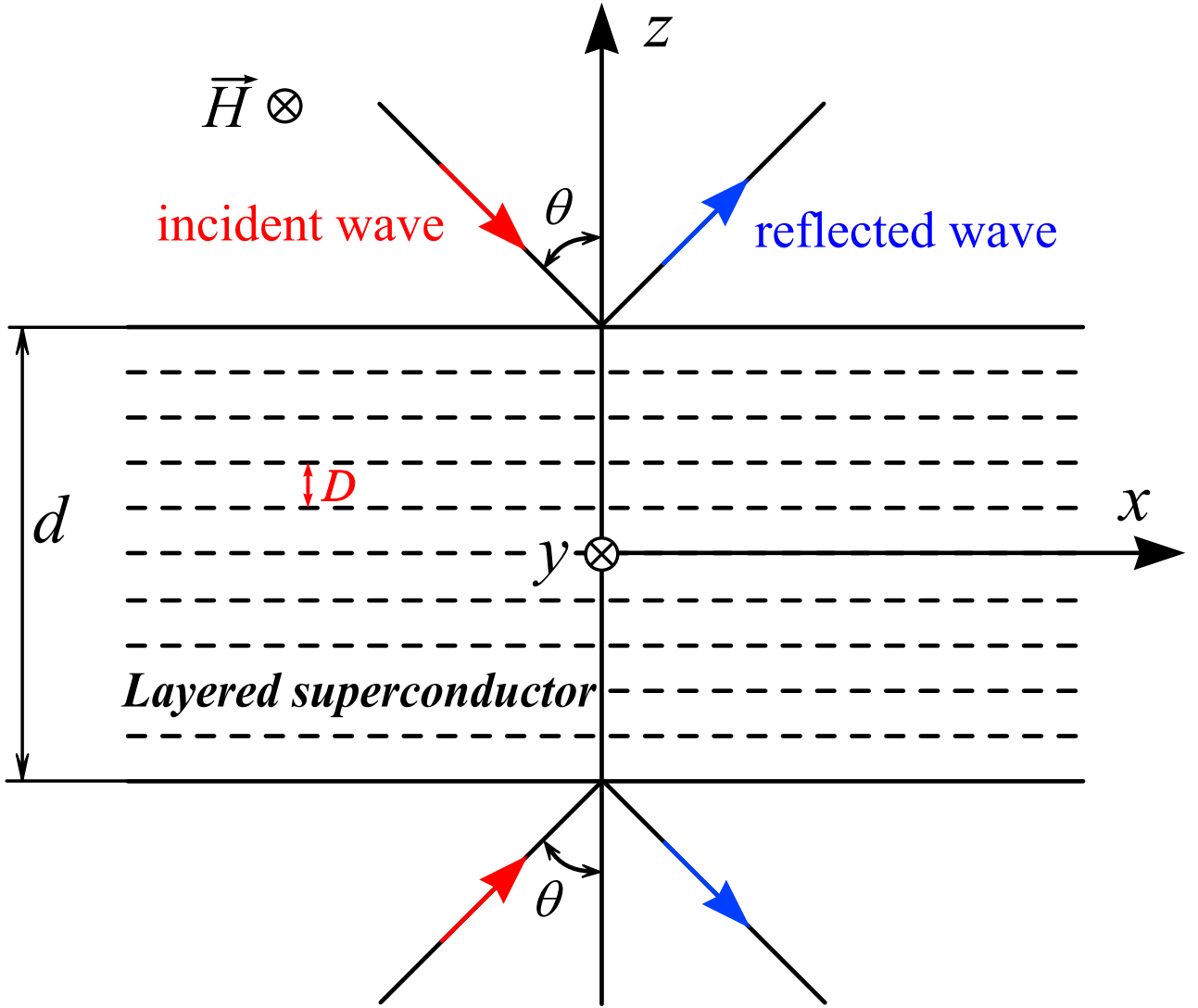}
\caption{(Color online) Geometry of the problem. The slab of a
layered superconductor is irradiated with a $p$-polarized
electromagnetic wave with the magnetic field symmetrical with
respect to  the middle of the sample.}\label{fig1}
\end{center}
\end{figure}

Consider a slab of a layered superconductor of thickness $d$ (see
Fig.~\ref{fig1}). The crystallographic \textbf{ab}-plane coincides
with the $xy$-plane and the \textbf{c}-axis is along the $z$-axis.
The interlayer distance $D$ is much smaller than the thickness $d$
of the slab.

Let the sample be irradiated by two $p$-polarized (Transverse
Magnetic) plane monochromatic electromagnetic waves with the
magnetic fields symmetric with respect to the middle of the
sample, the plane $z=0$. Therefore, the magnetic ${\vec H} = \{0,
H, 0\}$ and electric ${\vec E} = \{E_x, 0, E_z\}$ fields satisfy
the symmetry conditions,
\[
H(x,z,t)=H(x,-z,t),
\]
\begin{equation}\label{magn1}
 E_x(x,z,t)=-E_x(x,-z,t),\, E_z(x,z,t)=E_z(x,-z,t).
\end{equation}
Due to this symmetry, we will only consider the field distribution
in the upper half-space $z>0$.

The electromagnetic field in the vacuum, $z>d/2$, is the sum of
the incident and specularly reflected waves. The Maxwell equations
give for them,
\begin{equation}\label{24}
\begin{split}
H^V(x,z,t)=& H_0\cos\gamma_-+H_r\cos(\gamma_++\chi),\\
E_{x}^V(x,z,t)=&-\frac{k_z}{k}\Big(H_0\sin\gamma_--H_r\sin(\gamma_++\chi)\Big),\\
\gamma_{-}=k_xx -& k_z z-\omega t,\quad \gamma_{+}=k_xx + k_z
z-k_zd-\omega t,
\end{split}
\end{equation}
with $k_x=k\sin\theta$, $k_z=k\cos\theta$, $k=\omega/c$. Here
$\omega$ is the wave frequency, $c$ is the speed of light, and
$\theta$ is the angle of incidence. The value of $\chi$ is the
\emph{phase shift of the reflected wave} at the boundary $z=d/2$
of the slab. As is known, $\chi$ defines the \emph{surface
reactance} $X$ of a sample: when neglecting the dissipation,
\[
X=\frac{4\pi}{c}\tan\left( \frac{\chi}{2}\right)\cos\theta .
\]
%%%%%%%%%%%%%%%

Inside a layered superconductor, the electromagnetic field is
determined by the interlayer gauge-invariant phase difference
$\varphi$ of the order parameter. The spatial distribution of
$\varphi (x,z,t)$ obeys the set of coupled sine-Gordon equations
(see, e.g., Ref.~\onlinecite{bar,sine-gord}). We consider the
nonlinear JPWs with $|\varphi|\ll 1$, when the Josephson current
$J_c\sin\varphi$ can be approximated by $J_c
(\varphi-\varphi^3/6)$. We also assume that the gauge-invariant
phase difference experiences small changes on the scale $D$ in the
$z$-direction, and thus we can use the continuum approach. In the
continuum limit, the coupled sine-Gordon equation has the form,
\begin{equation}\label{e12}
\left(1-\lambda_{ab}^2\frac{\partial^2}{\partial
z^2}\right)\left(\frac{1}{\omega_J^2}\frac{\partial^2
\varphi}{\partial t^2} + \varphi-\frac{\varphi^3}{6}\right)-
\lambda_c^2\frac{\partial^2 \varphi}{\partial x^2}=0.
\end{equation}
Here $\lambda_{ab}$ and $\lambda_{c}=c/\omega_J\varepsilon^{1/2}$
are the London penetration depths across and along layers,
respectively, and $\omega_J = (8\pi e D
J_c/\hbar\varepsilon)^{1/2}$ is the Josephson plasma frequency.
The latter is determined by the maximum Josephson current $J_c$,
the interlayer dielectric constant $\varepsilon$, and  the
interlayer spacing $D$. The spatial variations in the
$z$-direction of the fields inside the very thin superconducting
layers are neglected. Here we also omit the dissipation terms
related to the quasiparticle conductivity. They are controlled by
the sample temperature $T$ and can be reduced to negligibly small
values. Moreover, in Eq.~(\ref{e12}), we neglect the term with the
capacitive coupling, for waves with sufficiently high
$k_x=k\sin\theta \sim \omega/c \gg \beta/\lambda_c$. Here $\beta =
R_D^2\varepsilon /sD \ll 1$ is the prefactor of the capacitive
coupling~\cite{koy}, $R_D$ is the Debye length for a charge in a
superconductor, and $s$ is the thickness of the superconducting
layers.

The magnetic and electric fields in a layered superconductor are
related to the gauge-invariant phase difference as
\begin{gather}
\frac{{\partial H ^S}}{{\partial x}} = \frac{{\cal H}_0}{\lambda
_c}\left( \frac{1}{{\omega_J ^2 }}\frac{{\partial ^2 \varphi
}}{{\partial t^2 }} + \varphi  - \frac{{\varphi ^3 }}{6}
\right),\notag\\
E_{x} ^S  = - \frac{{\lambda _{ab} ^2 }}{c}\frac{\partial^2 H ^S
}{\partial z \,\partial t}, \quad {\cal H}_0 =\frac{\Phi _0}{2\pi
D\lambda_c},\label{supconfield}
\end{gather}
where $\Phi_0=\pi c \hbar/e$ is the flux quantum and $e$ is the
elementary charge.

As was shown in Ref.~\onlinecite{n-lin}, the \emph{nonlinearity
in} Eq.~(\ref{e12}) \emph{can play a crucial role in the wave
propagation for frequencies close to} $\omega_J$, i.e., for
$|1-\Omega ^2|\equiv |1-\omega^2/\omega_J^2|\ll 1$. Indeed, if
$\varphi \sim |1-\Omega ^2|\ll 1$, \emph{the cubic term}
$\varphi^3$ in Eq.~(\ref{e12}) \emph{is of the same order as the
linear term} $\omega_J^{-2}\partial^2 \varphi/\partial
t^2+\varphi$.

We consider the frequency range below the Josephson plasma
frequency, $\Omega <1$, and seek a solution of Eq.~(\ref{e12}) of
the form,
\begin{equation}\label{solution}
\begin{split}
\varphi(x,z,t)=&a(z)(1-\Omega^2)^{1/2}\sin(\gamma_{0}+\alpha),
\\
\gamma_{0}=&k_xx -k_zd/2-\omega t,
\end{split}
\end{equation}
keeping only the first harmonics in $(k_x x-\omega t)$.

Substituting $\varphi$ in Eq.~(\ref{solution}) into
Eq.~(\ref{supconfield}), one obtains
\begin{equation}\label{H(z)}
\begin{split}
&H ^S(x,\zeta,t)=-{\cal H}_0  \frac{(1-\Omega^2)}{\kappa}h(\zeta)\cos(\gamma_{0}+\alpha),\\
&E_x ^S(x,\zeta,t)={\cal H}_0 \frac{(1-\Omega^2)}{\kappa}
P\,h'(\zeta)\sin(\gamma_{0}+\alpha).
\end{split}
\end{equation}
Here we introduce the dimensionless variables,
\begin{equation}\label{efh}
h(\zeta)=a(\zeta)-\frac{a^3(\zeta)}{8},\quad \zeta=\frac{\kappa
z}{\lambda_{ab}},
\end{equation}
and parameters,
\begin{equation}\label{efh1}
P=\frac{\lambda
_{ab}}{\lambda_c}\frac{\kappa}{\sqrt{\varepsilon}}, \quad
\kappa=\frac{\lambda_c k_x}{(1-\Omega^2)^{1/2}},
\end{equation}
and the prime denotes $d/d\zeta$.

Equations (\ref{e12}) and (\ref{solution}) yield the ordinary
second-order differential equation for $a(z)$:
\begin{equation}\label{a_xi}
\left[1-\kappa^2\displaystyle\frac{d^2}{d\xi^2}\right]
\left(a-\displaystyle\frac{a^3}{8}\right) +\kappa^2a=0.
\end{equation}
Further, we also assume $\kappa \gg 1$, which is valid for not
very small incident angles $\theta$. In this case, integrating
Eq.~(\ref{a_xi}) with the symmetry condition, $$a'(0)=0,$$ we
obtain
\begin{equation}\label{difeq}
\frac{3}{4}(a')^2=\left(\frac{8-3a_0^2}{8-3a^2}\right)^2-1,
\end{equation}
where $a_0=a(0)$. The solution of Eq.~(\ref{difeq})can be written
in the implicit form,
\begin{equation}\label{sol-xi}
\zeta=\sqrt{\frac{3}{4}}\int\limits_{a_0}^{a(\zeta)}da
\frac{8-3a^2}{\displaystyle\sqrt{(8-3a_0^2)^2-(8-3a^2)^2}}.
\end{equation}

\begin{figure}[!htp]
\begin{center}
\hspace*{-0.9cm}
%\hspace*{-0.5cm}\includegraphics*[width=8cm]{fig2.eps}
\hspace*{-0.5cm}\includegraphics*[width=17.5cm]{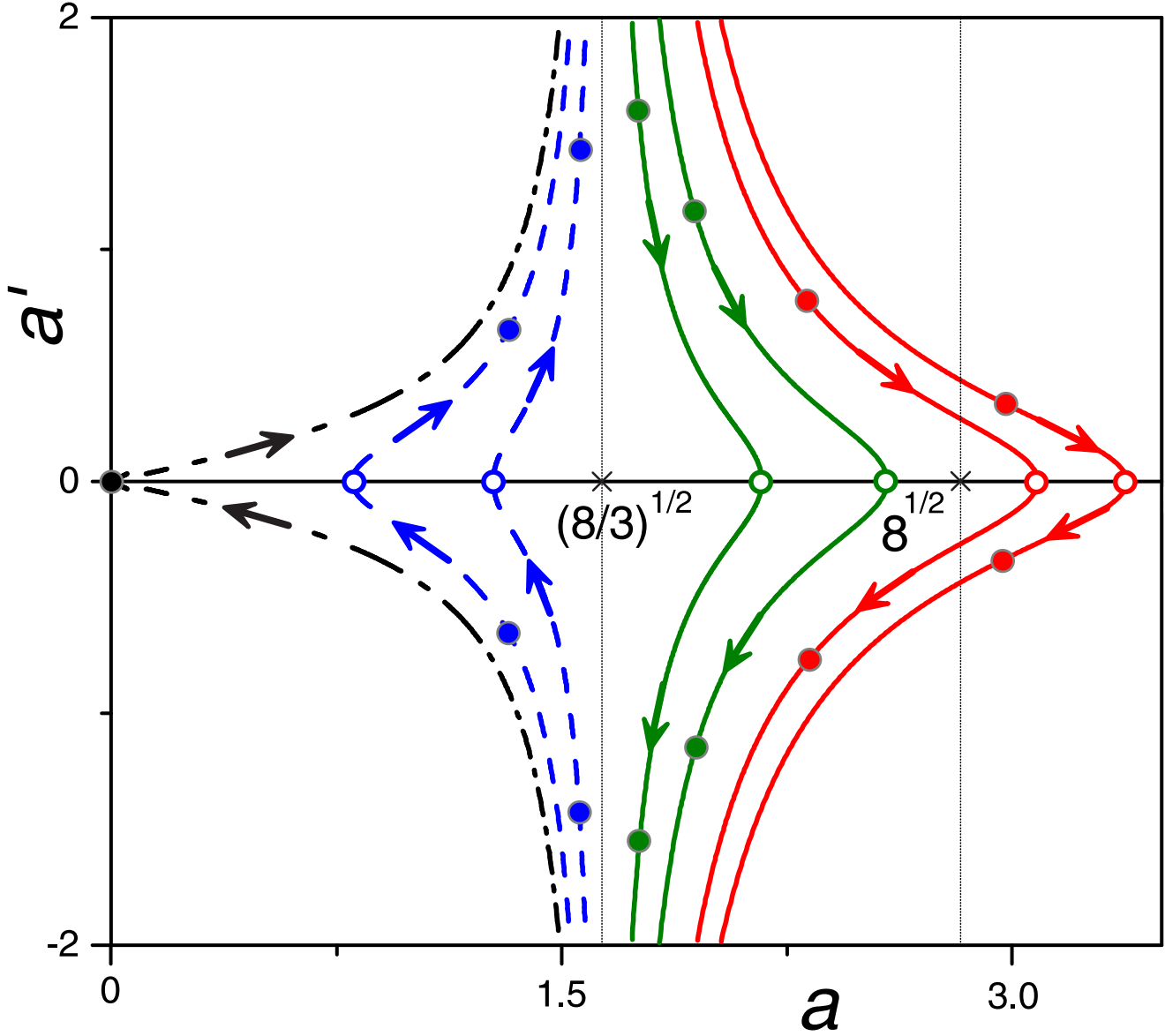}
\caption{(Color online) The phase diagram $a'(a)$ (for $a>0$).
Recall that $a$ is the amplitude of the gauge-invariant phase
$\varphi$ (see Eq.~(\ref{solution})), while $a'$ is its derivative
with respect to the dimensionless coordinate $\zeta = \kappa
z/\lambda_{ab}$. The paths along the arrows on the phase
trajectories between solid circles correspond to the change of the
coordinate $z$ inside the sample, from $z=-d/2$ to $z=d/2$. Open
circles correspond to the middle of the slab ($z=0$).}\label{fig2}
\end{center}
\end{figure}

The phase diagram, i.e., the set of $a'(a)$ curves for different
values of the constant $a_0$, is shown in Fig.~\ref{fig2}. Solid
circles mark the sample boundaries, while open circles indicate
the middle of the slab. Arrows show the direction of motion along
the phase trajectories when increasing the coordinate $\zeta$.

Thus the electromagnetic fields in the vacuum and in the
superconducting slab are determined by Eqs.~(\ref{24}) and
(\ref{H(z)}), respectively. The latter equations contain the
function $a(\zeta)$ given by Eq.~(\ref{sol-xi}).

\section{Phase shift of the reflected wave}

Now we find the relationship between the phase shift $\chi$ of the
reflected wave and the amplitude $H_0$ of the incident wave. With
this purpose, we join the tangential components of the electric
and magnetic fields in the vacuum and in the superconducting slab,
at the interface $z=d/2$. Thus, separating terms with $\sin(k_x x
- \omega t)$ and $\cos(k_x x - \omega t)$, we derive four
equations for $H_r$, $H(d/2)$, $\chi$, and $\alpha$:
\begin{equation}\label{sys1}
\begin{split}
-h_0+h_r\cos\chi=&Ph'(\delta)\sin\alpha,\\
-h_0-h_r\cos\chi=&h(\delta)\cos\alpha,\\
-h_r\sin\chi=&Ph'(\delta)\cos\alpha,\\
-h_r\sin\chi=& h(\delta)\sin\alpha.
\end{split}
\end{equation}
Here
\[
h_0=\frac{H_0}{{\cal H}_0 } \frac{\kappa}{(1-\Omega^2)}, \quad
h_r=\frac{H_r}{{\cal H}_0 } \frac{\kappa}{(1-\Omega^2)};
\]
$\delta=\kappa d/2\lambda _{ab}$ is the value of $\zeta$ at the
interface $z=d/2$. Note that, without any loss of generality, we
can assume the value $h(\delta)$ of the total magnetic field at
the sample surface to be positive, $h(\delta)>0$. A negative
$h(\delta)$ correspond to replacing $\chi \rightarrow \chi +\pi$
in Eqs.~(\ref{sys1}).

Excluding $h_r$ and $\alpha$ from Eqs.~(\ref{sys1}), we find
\begin{equation}\label{AB4}
h_0=\frac{1}{2}\sqrt{h^2(\delta)+\big(P h'(\delta)\big)^2}
\end{equation}
\begin{equation}\label{AB5}
\chi=2\arctan{\left(\frac{P h'(\delta)}{h(\delta)}\right)}.
\end{equation}
Equations (\ref{AB4}), (\ref{AB5}), together with Eqs.~(\ref{efh})
and (\ref{sol-xi}), give, in an implicit form, the required
dependence of the phase shift $\chi$ on the amplitude of the
incident wave $H_0$.

\begin{figure}\centering
%\scalebox{1}[1]{\includegraphics[70,70][534,512]{phase.eps}}
%\includegraphics[width=8cm]{ha.eps}
\includegraphics[width=17cm]{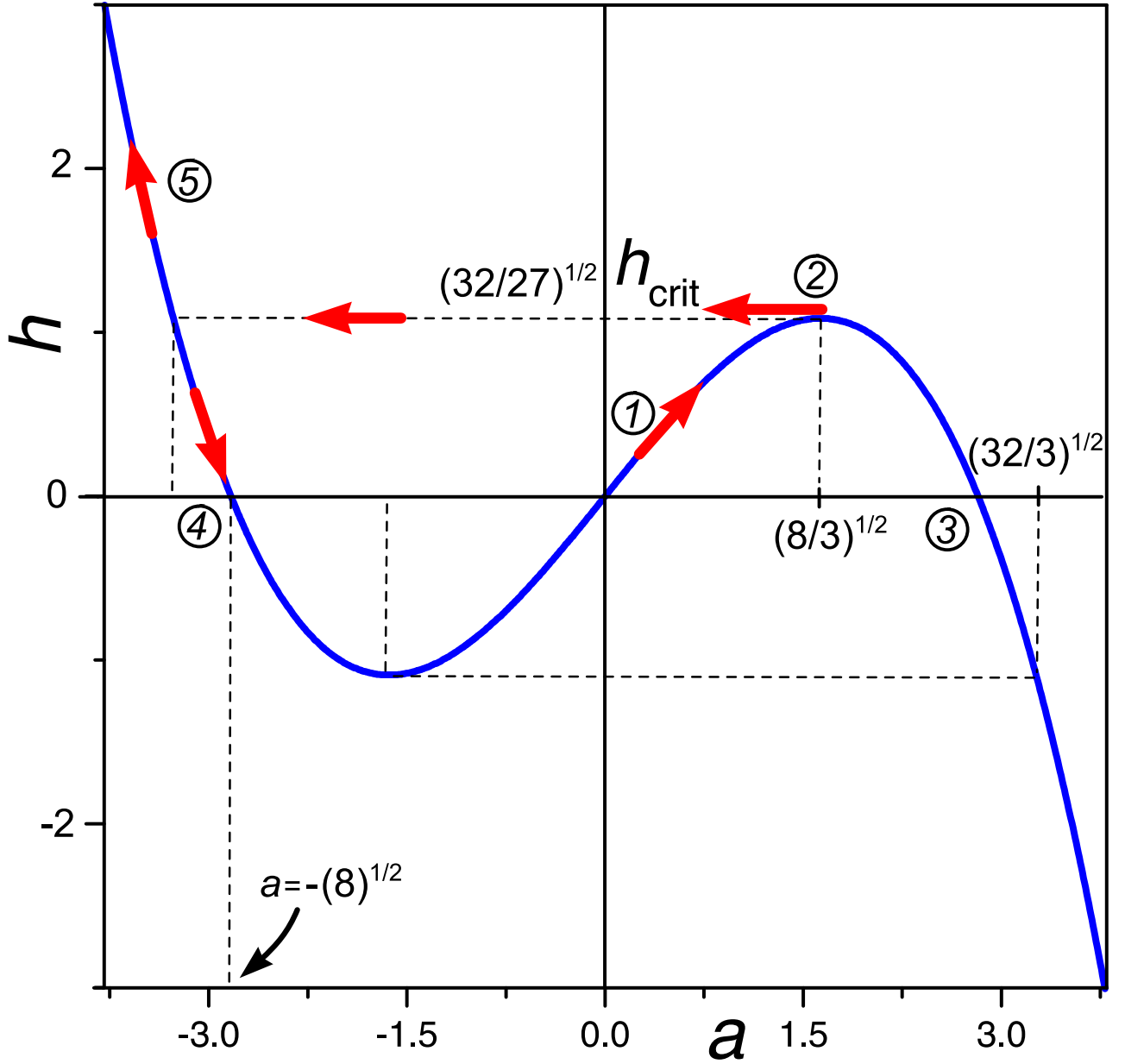}
\vspace*{-0.2cm} \caption{(Color online) The normalized magnetic
field $h$ versus the normalized amplitude $a$ of the
gauge-invariant phase $\phi$. This $h(a)$ dependence is described
by Eq.~(\ref{efh}). Arrows indicate the motion along the $h(a)$
curve when changing the normalized amplitude $h_0$ of the incident
wave.}\label{fig9}
\newpage
\end{figure}

For further analysis of the amplitude dependence of the phase
shift $\chi$, it is very important to take into account the
non-single-valued relation between the values $h(\delta)$ and
$a(\delta)$. Indeed, Eq.~(\ref{efh}) and Fig.~\ref{fig9} show that
there exist \emph{three values }of $a$ that correspond to the
\emph{same value} of $h$, if $0<h<(32/27)^{1/2}$. Taking into
account Eq.~(\ref{AB4}), we conclude that three different values
of the parameter $h'(\delta)/h(\delta)$ in Eq.~(\ref{AB5})
correspond to the same value of $h_0$. This results in the
appearance of \emph{three branches} of the dependence $\chi(h_0)$.

Below, for simplicity, we restrict ourselves to the case of small
sample thicknesses, when
\begin{equation}\label{d1}
\delta \ll 1.
\end{equation}
In this case, $|a(\zeta)-a_0| \ll 1$, and Eq.~(\ref{sol-xi}) is
significantly simplified,
\begin{equation}\label{th_a_impl}
 a(\zeta) \simeq a_0\Big(1+\frac{4 \zeta^2}{8-3a_0^2}\Big),
\end{equation}
and one can easily derive the asymptotic equations for all three
branches $\chi(h_0)$.

\subsection{Low-amplitude branch $\chi(h_0)$}

First, we discuss the branch of the dependence $\chi(h_0)$ that
corresponds to the portion ``1--2'' of the $h(a)$ curve in Fig.~3.
This monotonically increasing branch is shown by the blue curve
(with number 1) in the main panel of Fig.~4, and also in inset (a)
of Fig.~4. This blue branch is defined within the interval
$(0,h_{0\,\mathrm{cr}})$ of the $h_0$ change. The ending point
$h_{0\,\mathrm{cr}}$ of the branch is described by the expression,
\begin{equation}\label{ending}
h_{0\,\mathrm{cr}} \simeq
\sqrt{\frac{8}{27}}+\sqrt{\frac{3}{8}}P^2\delta^2.
\end{equation}
At this point,
\begin{equation}\label{chi-ending}
\chi\left(h_{0\,\mathrm{cr}}\right) \simeq
3P\delta-\sqrt{\frac{9}{8}}P\delta^2.
\end{equation}
\begin{figure}\centering
%\scalebox{1}[1]{\includegraphics[70,70][534,512]{phase.eps}}
\hspace*{-1.5cm} %\vspace*{-1cm}
\includegraphics[width=17cm]{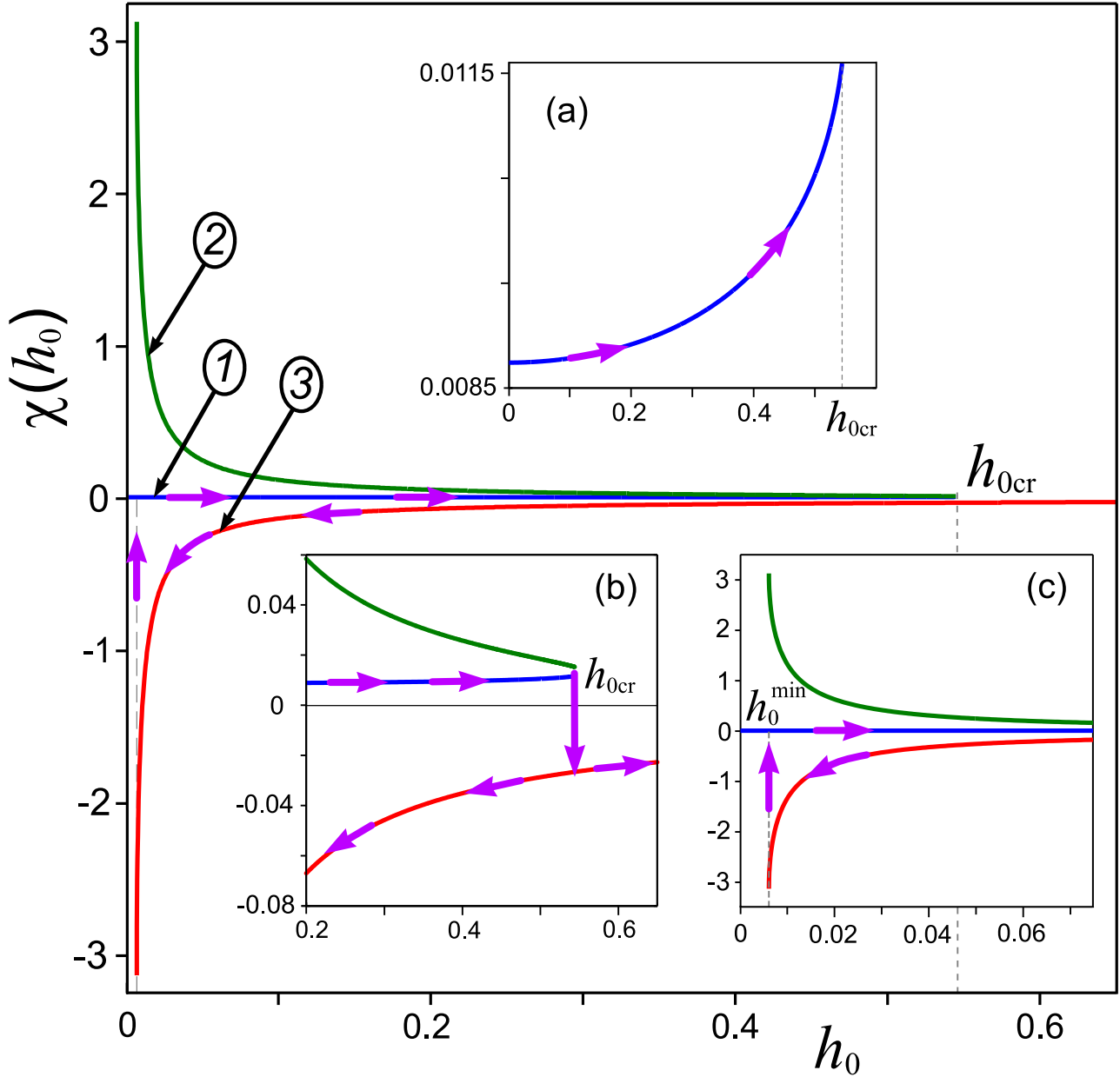}
\vspace*{0.015cm} \caption{(Color online) The numerically-obtained
dependence of the phase shift $\chi$ of the reflected wave on the
dimensionless amplitude $h_0$ of the incident electromagnetic
wave. The values of the parameters here are:
$d/\lambda_{ab}=0.05$, $\lambda_c/\lambda_{ab}=200$, $\kappa =10$,
and $\varepsilon=16$. The insets show magnified portions of the
$\chi(h_0)$ plots. The arrows on the curves indicate how the phase
shift $\chi$ changes when periodically varying the amplitude
$h_0$. The first jump in $\chi(h_0)$ at $h_0=h_{0 {\rm cr}}$ is
shown by the downwards vertical arrow in the inset (b). This
produces a jump $\Delta \chi \approx -9P\delta$. The second
(reverse) jump at $h_0=h_0^{\rm min}$ is shown in the main frame
and in the inst (c). Namely, the upward vertical arrow (from
$\left. \chi(h_0)\right|_{{\rm before}\,\,{\rm jump}}=-\pi$, to
$\left. \chi(h_0)\right|_{{\rm after}\,\,{\rm jump}} \approx 0$)
shows the jump in $\chi(h_0)$.}\label{chi_h0}
%\vspace*{10.5cm}
\end{figure}

The low-amplitude asymptotics of $\chi (h_0)$ dependence is,
\begin{equation}\label{low-am}
\chi \simeq 2P\delta\left(1+\frac{h_0^2}{2}\right), \quad h_0 \ll
1.
\end{equation}

\subsection{High-amplitude branch $\chi(h_0)$}
The second and third branches of the $\chi(h_0)$ dependence
correspond to the portions ``2--3'' and ``4--5'', respectively, of
the $h(a)$ curve in Fig.~3. The second and third branches of
$\chi(h_0)$  are shown by the green and red curves (indicated by
numbers 2 and 3, respectively) in Fig.~4. The second branch exists
in the interval
\begin{equation}\label{interv-2}
h_0^{\rm min} <h_0<h_{0\,\mathrm{cr}}, \quad h_0^{\rm min}
\simeq\sqrt{2}P\delta \,.
\end{equation}
As is seen in Fig.~4, the first and second branches of $\chi(h_0)$
almost meet at the point $h_{0\,\mathrm{cr}}$. The third branch is
defined for
\begin{equation}\label{interv-3}
h_0^{\rm min}<h_0<\infty.
\end{equation}

Near the minimum value of $h_0$, i.e., for $h_0 \sim h_0^{\rm
min}$, the second and third branches have the asymptotics,
\begin{equation}\label{chisq8}
\chi(h_0)=\pm\left(\pi-\sqrt{\frac{2h_0^2}{\delta^2}-4P^2}\right).
\end{equation}
Here, the signs ``+'' and ``$-$'' correspond to the second and
third branches, respectively.

The third branch tends to zero for $h_0 \rightarrow \infty$
following the expression,
\begin{equation}\label{infin}
\chi(h_0)=-P\delta\Big(\frac{4}{h_0}\Big)^{2/3}, \quad h_0\gg1 \,.
\end{equation}

Now we present the parametrically-defined formula for all three
branches, which is valid for $h_0$ not very close to
$h_{0\,\mathrm{cr}}$:
\begin{subequations}\label{chih0param}
\begin{eqnarray}
h_0(a_0)&=&\frac{a_0|8-a_0^2|}{16},
\label{h0param}\\
\chi(a_0)&=&\frac{16P}{8-a_0^2}\delta. \label{chiparam}
\end{eqnarray}
\end{subequations}
Figure~5 demonstrates the very good agreement of this formula with
numerical results.
\begin{figure}[!htp]\centering
%\scalebox{1}[1]{\includegraphics[70,70][534,512]{phase.eps}}
\hspace*{-1.5cm} \vspace*{-0.1cm}
\includegraphics[width=17cm]{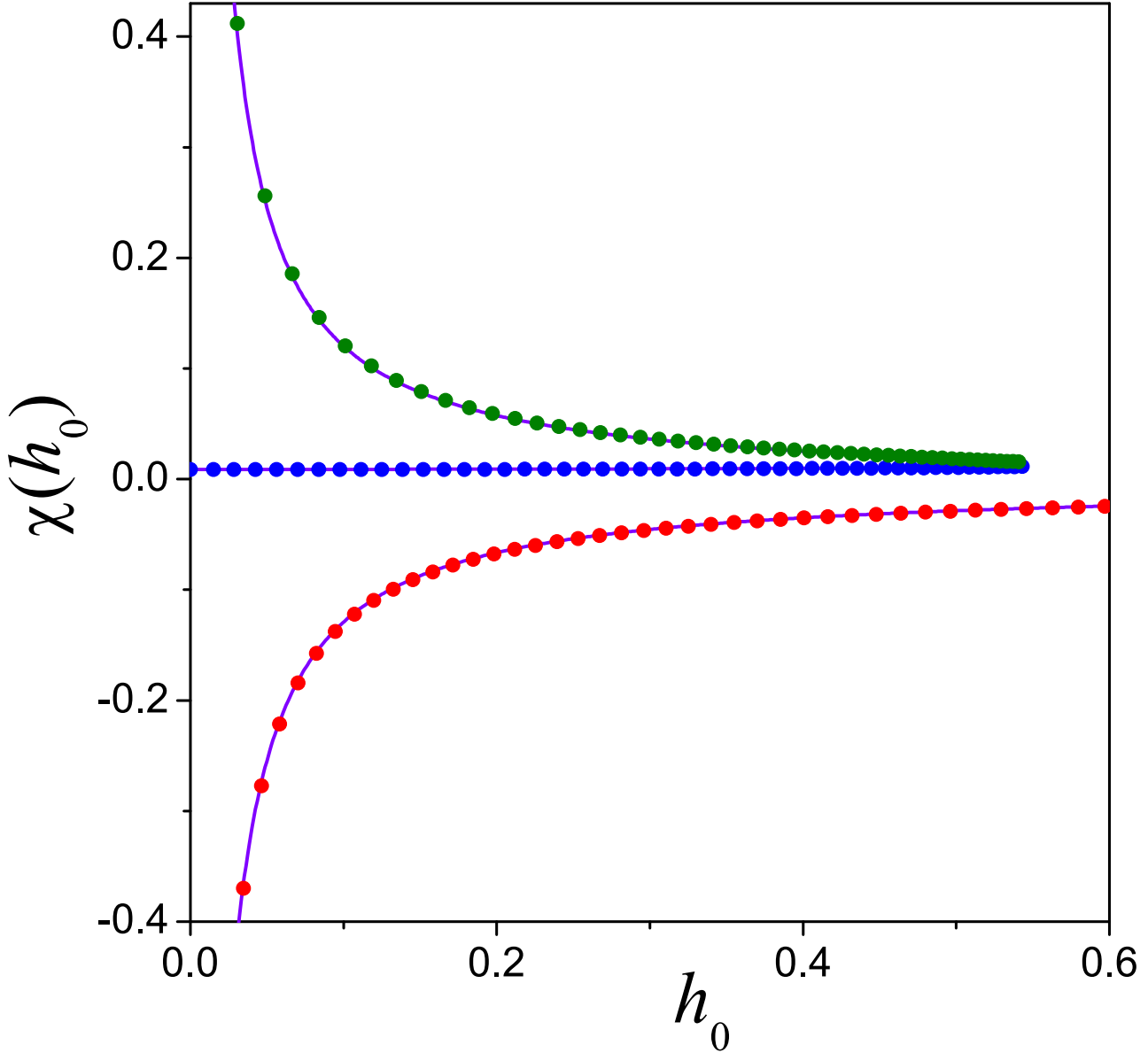}
\vspace*{-0.15cm} \caption{(Color online) The phase shift $\chi$
of the reflected wave versus the dimensionless amplitude $h_0$ of
the incident wave, for $d/\lambda_{ab}=0.05$,
$\lambda_c/\lambda_{ab}=200$, $\kappa=10$, and $\varepsilon=16$.
The solid curves are the plots of the three branches of the
function given by Eqs.~(\ref{chih0param}), while the dots were
numerically obtained from Eqs.~(\ref{efh}), (\ref{sol-xi}),
(\ref{AB4}), and (\ref{AB5}).}\label{chi_h02}
\newpage
\end{figure}

\subsection{Hysteresis jumps in the dependence $\chi(h_0)$}
From the analysis shown above, we can now describe the behavior of
the phase shift $\chi$ of the reflected wave when periodically
changing the amplitude $h_0$ of the incident wave.

When increasing $h_0$ from zero, the phase shift $\chi (h_0)$
increases monotonically following the first branch in Fig.~4. At
the point $h_0=h_{0\,\mathrm{cr}}$, the first branch comes to the
end, and a jump to the third branch should occur when further
increasing $h_0$. The phase shift $\chi (h_0)$ jumps from the
value $$\chi (h_{0\,\mathrm{cr}} - 0) \simeq 3P\delta$$ to the
final value $$\chi (h_{0\,\mathrm{cr}} + 0) \simeq -6P\delta .$$
Thus, the jump here is
\begin{equation}\label{1-jump}
\left. \Delta \chi \right|_{h_0=h_{0 {\rm cr}}}  \approx
-9P\delta.
\end{equation}

When decreasing $h_0$, the phase shift $\chi (h_0)$ decreases
following the third branch, crosses the point
$h_0=h_{0\,\mathrm{cr}}$, and only at the point
$h_0=h_0^{\mathrm{min}}$ it performs a reverse jump to the first
branch. Therefore, now
\begin{equation}\label{2-jump}
\left. \Delta \chi \right|_{h_0=h_0^{\mathrm{min}}}  \approx \pi.
\end{equation}

Thus, the hysteretic jumps $\Delta \chi $ of the phase $\chi
(h_0)$ of the reflected wave could be observed when periodically
changing the amplitude $h_0$ of the incident wave.

\section{Conclusion}
In this paper, we have predicted and theoretically analyzed
unusual phenomenon for conducting media. We have shown that, due
to the specific nonlinearity of layered superconductors,
\emph{hysteretic jumps of the surface reactance} could be observed
when periodically changing the amplitude $H_0$ of the incident
wave. A remarkable feature of the predicted phenomenon is the
\emph{relatively small values of the necessary ac amplitudes}
$H_0$. The hysteretic jumps can be observed \emph{even when}
$\varphi \ll 1$. According to our analysis, the critical amplitude
$$H_{0\,\mathrm{cr}} =h_{0\,\mathrm{cr}}\cdot{\cal H}_0
\frac{(1-\Omega^2)}{\kappa} \ll {\cal H}_0 \sim 20\,\,{\rm Oe}.$$
There are two small parameter here, $(1-\Omega ^2)$ and
$1/\kappa$. As was shown in Ref.~\onlinecite{n-lin}, the heating
effect is negligible for such ac amplitudes.

The phenomenon discussed here is another exciting example of the
numerous unusual effects related to the very specific nonlinearity
of layered superconductors.

\begin{acknowledgments}

We gratefully acknowledge partial support from the National
Security Agency (NSA), Laboratory of Physical Sciences (LPS), Army
Research Office (ARO), National Science Foundation (NSF) grant No.
EIA-0130383, JSPS-RFBR 06-02-91200, and Core-to-Core (CTC) program
supported by Japan Society for Promotion of Science (JSPS). S.S.
acknowledges support from the Ministry of Science, Culture and
Sport of Japan via the Grant-in Aid for Young Scientists No
18740224, the EPSRC via No. EP/D072581/1, EP/F005482/1, and ESF
network-programme ``Arrays of Quantum Dots and Josephson
Junctions''.

\end{acknowledgments}

%\vspace{-0.5cm}

\end{document}